\documentclass[preprint2]{aastex}

\newcommand{\etal}{{\it et al.\,}}
\newcommand{\teff}{T$_{\rm eff}$\,}
\newcommand{\logg}{log~g\,}

\shorttitle{UMPS and dust-gas separation}
\shortauthors{Venn \& Lambert}

\begin{document}


\title{Could the Ultra Metal-poor Stars be Chemically Peculiar
  and Not Related to the First Stars?}


\author{K. A. Venn}
\affil{Department of Physics \& Astronomy, University of Victoria,
       Elliott Bldg, 3800 Finnerty Road, Victoria, BC, V8P 5C2, Canada}
\email{kvenn@uvic.ca}
\and
\author{David L. Lambert}
\affil{The W.J. McDonald Observatory, The University of Texas at Austin,
       RLM 15.308, Austin, TX, 78712}
\email{dll@astro.as.utexas.edu}


\begin{abstract}

Chemically peculiar stars define a class of stars that show
unusual elemental abundances due to stellar photospheric effects
and not due to natal variations. 
In this paper, we compare the elemental abundance patterns of the
ultra metal-poor stars with metallicities [Fe/H] $\sim -5 $ to 
those of a subclass of chemically peculiar stars.   These
include post-AGB stars, RV Tauri variable stars, and the Lambda Bootis 
stars, which range in mass, age, binarity, and evolutionary status, yet 
can have iron abundance determinations as low as [Fe/H] $\sim -5$.
These chemical peculiarities are interpreted as due to the separation 
of gas and dust beyond the stellar surface, followed by the accretion 
of dust depleted-gas.  
Contrary to this,
the elemental abundances in the ultra metal-poor stars are thought  
to represent yields of the most metal-poor supernova
and, therefore, observationally constrain the earliest stages of
chemical evolution in the Universe.  
Detailed chemical abundances are now available for HE1327-2326 
and HE0107-5240, the two extreme ultra metal-poor stars
in our Galaxy, and for HE0557-4840, another ultra metal-poor star 
found by the Hamberg/ESO survey.  
There are interesting similarities in their abundance ratios 
to those of the chemically peculiar stars, 
e.g., the abundance of the elements in their photospheres are
related to the condensation temperature of that element. 
If HE1327-2326 and HE0107-5240 are ultra metal-poor due 
to the preferential removal of metals by dust grain formation or
dilution through the accretion of metal-poor interstellar gas,
then their CNO abundances suggest true metallicities of [$X$/H] $\sim -2$
rather than their present metallicities of [Fe/H] $\leq -5$,
and, thus their status as truly ultra metal-poor stars is called into question.
The initial abundance for HE0557-4840 would be [$X$/H] $\ge -4$.  
It is important to establish the nature of these stars since they are
used as tests of the early chemical evolution of the Galaxy, yet if 
they are chemically peculiar, then those tests should be focused on 
stars in the metal-poor tail of the Galactic halo distribution.

Many, but not all, chemically peculiar stars show a mid-infrared 
excess from the circumstellar dust.   We examine the JHK fluxes for 
the ultra metal-poor stars but find no excesses.   
A more important test of the stars' status as
chemically peculiar is provided by the elemental abundances 
of sulphur and/or zinc. These two elements have low condensation 
temperatures and do not form dust grains easily; furthermore, 
the chemically peculiar stars universally show sulphur and zinc 
to be undepleted or nearly so. 
We show that near-infrared lines of S\,{\sc i} offer a promising test of 
the possibility that HE1327-2326 may be chemically peculiar.
Although there are some parallels between the compositions of the
ultra metal-poor stars and chemically peculiar stars, a definitive 
ruling on whether the former are chemically peculiar requires additional
information.

\end{abstract}




\section{Introduction}

A fossil record of the earliest episodes of stellar nucleosynthesis 
in the Universe and Galaxy should be revealed by the compositions of 
the most metal-poor Galactic stars
(e.g., Tumlinson 2007a, 2007b; Tominaga \etal 2007; Umeda \& Nomoto 2003). 
The lure of this revelation
has driven the search to find and analyse such Rosetta stones. 
A great leap forward  was achieved recently by the 
discovery of two stars with  iron abundances [Fe/H] $< -5.3$
(Christlieb \etal 2002; Frebel \etal 2005),
a limit about 1.5 dex below the abundance of the previously known 
most metal-poor star. A third star with [Fe/H] $\simeq -4.8$ also
beats the previous lower bound (Norris \etal 2007). 
Prior to these remarkable discoveries, the most Fe-poor stars known  
were HR~4049 and HD~52961 with [Fe/H] $\simeq -4.8$ 
(e.g., Waelkens \etal 1991).   However, these and slightly more 
iron-rich examples were dismissed - correctly - as irrelevant 
to the issue of early stellar nucleosynthesis because they are 
`chemically peculiar', i.e., their present surface compositions 
are far removed from their initial compositions.
In particular, their compositions reflect that of gas from which
refractory elements have been removed to varying degrees by a process
dubbed `dust-gas separation'.
The existence of HR~4049 and HD~52961 has led us to reexamine the 
question of whether the recently discovered ultra metal-poor
stars may themselves be chemically peculiar. 

Two of the three ultra metal-poor stars in question are 
HE1327-2326 and HE0107-5240.
HE1327-2326 was discovered by Frebel \etal (2005) and abundance analyses
have been described by Aoki \etal (2006), Frebel \etal (2006), 
Collet \etal (2006), and most recently by Frebel \& Christlieb (2007). 
The latter analysis based on
the highest SNR spectra yields abundances for 11 elements and
upper limits for an additional nine elements: the new iron abundance
at [Fe/H] $=-5.9$
is even lower than the previous determination.
HE0107-5240 was discovered by Christlieb \etal (2002) and abundance
analyses have been reported by Christlieb \etal (2004), Bessell \etal 
(2004), and Collet \etal (2006) with the latter suggesting
[Fe/H]$\simeq -5.6$. 
The third star HE0557-4840 was discovered and analysed by 
Norris \etal (2007): [Fe/H] $= -4.8$ places it between 
the two ultra metal-poor stars (HE1327-2326 and HE0107-5240) 
and the lower boundary of the metal-poor tail of Galactic stars. 

A marked characteristic of these three discoveries is that some
abundance ratios are uncharacteristic  of metal-poor stars
of higher Fe abundance. Notably, the stars are C-rich for their
Fe abundance, i.e., [C/Fe] $= 3.7$ for HE1327-2326, and also
[N/Fe] $= 4.1$ and [O/Fe] $= 3.4$ (Frebel \& Christlieb 2007). This
property of unusual abundance ratios is shared in a qualitative sense with 
HR~4049 and HD~52961. 

In the following sections, we briefly review the classes of known 
chemically peculiar stars affected by dust-gas separation. 
Then, we discuss if the three stars HE1327-2326,  HE0107-5240, 
and HE0557-4840 are chemically peculiar rather than true 
ultra metal-poor stars. 
In the final sections, we discuss possible tests of the hypothesis 
that ultra metal-poor stars may be chemically peculiar.


\section {Separation of gas and dust and chemically peculiar stars}

The chemically peculiar stars in question are those whose
atmospheres betray the operation of the dust-gas separation process. 
In gas of a sufficiently low temperature, dust condenses out and 
the gas is depleted in those elements that form the dust. 
The local interstellar gas, for example, displays such depletion 
(Savage \& Sembach 1996). 
The composition of the gas in a dust-gas mixture depends on several 
primary factors including the initial composition of the gas, the total
pressure, and the  history of the gas-dust mixture. 
If a star were to then accrete gas, preferentially over dust,
and the accreted gas were to comprise a major fraction of the
stellar photosphere,  a star with striking abundance anomalies 
results.   This scenario is one version of how a star could develop
chemical peculiarities from dust-gas separation (or ``dust-gas winnowing'').
Anomalies plausibly attributable to dust-gas separation have now been
reported for three kinds of stars: Lambda Bootis stars,
post-AGB A-type spectroscopic binaries, and RV Tauri variables. 

The Lambda Bootis stars are main sequence stars, ranging from A-
to mid F-types, and found at all evolutionary phases from very young 
(e.g., they are found in young open clusters, Gray \& Corbally 1998), 
to the end of the main-sequence (Kamp \& Paunzen 2002, 
Andrievsky \etal 2002). 
They are expected to have a solar-like composition, but have 
long been known for their significant metal-deficiencies 
(Morgan, Keenan, \& Kellman 1943; Burbidge \& Burbidge 1956; 
Baschek \& Searle 1969).  These stars have effective temperatures 
of from about 6500 to 9500~K and surface gravities \logg $\simeq 4$. 
Our (Venn \& Lambert 1990) abundance analysis of three stars, 
including the eponym, showed that the compositions of the stars 
could be attributed to accretion of circumstellar gas, without dust.
The iron deficiency in the case of $\lambda$~Boo was [Fe/H] $\simeq -2$, 
and slightly less severe for 29~Cyg, with normal abundances of 
C, N, O and S for both stars, in agreement with the expectations
for an atmosphere contaminated with  dust-free gas (see Figure~1).  
Even Vega, a `standard' A0 star, was shown to be a mild Lambda Bootis 
star (Venn \& Lambert 1990, Lemke \& Venn 1996).   Vega also shows an
infrared excess due to a dusty circumstellar disk (Su \etal 2005;
Aumann \etal 1984).
  
The shallow convective envelope of early A-type stars is 
a key factor in the creation and maintenance of the abundance 
anomalies in the {\it diffusion/accretion model} 
(Turcotte \& Charbonneau 1993).
Abundance anomalies can persist ($\sim$10$^6$ yr), 
even after dispersal of the circumstellar
dust and gas and, hence, removal of the infrared excess contributed by 
the dust.   Thus, not all Lambda Bootis stars show an infrared excess; 
Paunzen \etal (2003) estimate that 23\% of {\it bona fide} Lambda Bootis 
stars show evidence for circumstellar matter.
On the other hand, the circumstellar matter may be of interstellar 
origin (Kamp \& Paunzen 2002, G\'{a}sp\'{a}r \etal 2007).   
Movement of a star through the denser parts of the interstellar medium 
can create a bow shock which heats the interstellar dust causing an 
infrared excess.  Meanwhile, radiation pressure from the star repels the 
grains, while gas is accreted onto the stellar surface.
With return of the star to passage through low density gas, the 
infrared excess dissipates and accretion ceases. 
This alternate theory for the origin of the Lambda Bootis stars 
implies the chemical anomalies are transient, but could help to
explain why the phenomenon is seen in such a wide range of 
main-sequence stars.

Abundance anomalies should not survive the transition from the 
main sequence to the giant branch though.  The deep convective 
envelope of giant stars will surely dilute abundance anomalies 
beyond recognition.   
Yet, a metal deficiency of even greater severity can be found among 
post-AGB stars in spectroscopic binaries (Waelkens \etal 1991;
Van Winckel 2003). These stars, like HR~4049 and HD~52961 discussed 
above, are supergiants with \teff in the range of 6000 to 7600~K and 
surface gravities \logg $\simeq 1$. One must suppose that a new
episode of dust-gas separation led to these abundance anomalies.

The original quartet of post-AGB binaries (Van Winckel \etal 1995)
comprised HR~4049, HD~44179, HD~52961, and BD~+39~4926.  Their [Fe/H] 
values range from $-3.0$ to $-4.8$, but they show an abundance pattern 
reminiscent of interstellar gas, i.e., quasi-solar abundances 
of C, N, O, S, and Zn, but severe underabundances of, for example, 
Si, Ca, and Fe (see Figure~1). 
The pattern shows that it is the former set of elements that define the
initial composition of these stars and not the latter set.
Gas is thought to be accreted onto the star from a circumbinary
disk, while radiation pressure exerted by the star on the dust grains
inhibits the accretion of dust by the star and may also promote a 
separation of dust and gas in the disk.   The dusty disk is betrayed 
by an infrared excess:
HD~44179, also known as the Red Rectangle, is a rather special 
proto-planetary nebula with a striking infrared excess.  On the other hand, 
BD~+39~4926 lacks an infrared excess.   Shallow convective envelopes in 
these extended stars are presumably a key factor in the appearance 
of their huge abundance anomalies.  

Chemical peculiarities of the post-AGB stars are presumably developed 
from less extreme peculiarities seen in their immediate progenitors, 
the RV Tauri variables.  Discovery of this third category of star 
displaying the marks of dust-gas winnowing began with the analysis 
of the RV Tauri variable IW~Car (Giridhar \etal 1994). 
A RV Tauri star is a post-AGB star with an infrared excess 
(first noted by Gehrz 1972).  
Subsequent analyses (Giridhar \etal 2005) showed that the effects of 
dust-gas separation among RV Tauri stars appear limited to the warmer 
stars (\teff $> 4500$ K) and to the stars with intrinsic metallicities 
[Fe/H] $\geq -1$.   Affected stars have \teff $\simeq 4500 - 6500$~K 
(hotter stars fall outside the instability strip and appear as 
non-variable post-AGB stars) and \logg $\simeq 0$ to 1. 
Gonzalez \& Lambert (1997) also discussed the potential importance of
the composition of the photosphere (e.g., C/O ratio), and environment
(e.g., field vs globular cluster stars).

 The reasons for
these effective temperature and metallicity
 boundaries are not entirely clear, but the cooler stars possess more
extensive convective envelopes that dilute the accreted gas.  
Dust-gas winnowing may be impaired in low metallicity gas where the dust 
to gas ratio is necessarily lower. 
The winnowing site is again presumed to be a circumbinary disk; 
there is increasing evidence that the affected stars are spectroscopic 
binaries (Van Winckel 2007).  How the gas is captured onto the star 
from the circumbinary disk in the presence of a stellar wind remains 
an unsolved problem, as it does for the post-AGB stars in binaries.

A signature of a star that is affected seriously by dust-gas winnowing 
is a correlation between an element's abundance and the predicted
condensation temperature $T_{cond}$ (or the abundance in interstellar gas).
Estimates of $T_{cond}$ depend on the  initial composition and pressure
in a gas and the assumption that cooling of the gas and condensation
of grains occurs under equilibrium conditions. Lodders (2003) provides
a comprehensive discussion of $T_{cond}$ estimates for all elements
for the solar (O-rich) composition. In the post-AGB stars, and in some 
Lambda Bootis and RV Tauri stars, the
abundance [$X$/H] is smoothly correlated with $T_{cond}$ (see Figure~1).

%

In some Lambda Bootis and RV Tauri stars, 
[$X$/H]  shows no obvious trend with
the  $T_{cond}$ (e.g., EQ~Cas, Giridhar \etal 2005).
Rao \& Reddy (2005) brought order to chaos by noting instead 
that the [$X$/H] for EQ~Cas were correlated with the ionization potential 
of the neutral atom (`the first ionization potential' or FIP). 
This raises the intriguing possibility that an alternative
or additional process is operating in this one star. 
The FIP effect is a well known  phenomena in the solar corona 
thought to reflect the greater ease with which 
ions (low FIP elements) rather than neutral atoms (high FIP elements) 
are fed from the cool chromosphere into the corona.
Perhaps, EQ~Cas is a star where there is a selective
feeding of the stellar wind. Thus,  
in single stars, the {\sl stellar wind} may control the abundance 
anomalies but where the variable is in a  binary the {\sl circumbinary
disk} may exert control. 

Another possibility for affecting the [$X$/H] vs $T_{cond}$ correlation 
(at high T$_{cond}$) is a competition between accretion of metal free
gas and chemical separation due to, e.g., diffusion, gravitational
settling, radiative acceleration, or rotational mixing above the
convective zone.   The diffusion/accretion model for Lambda Bootis stars
by Turcotte \& Charbonneau (1993) suggests timescales for chemical
separation can vary by element, thus complicating the abundance
pattern established earlier by accretion.
The conclusion must be that the dust-gas separation mechanism
does not have the same effects on the elemental abundances in
the various stars where it appears to operate.  
Either the mechanism itself, and/or the re-accretion phase of 
the dust-depleted gas, and/or additional processes that affect
chemical separation impact the observed chemical abundance pattern.

In summary, 
stars with abundance anomalies attributable to dust-gas separation
are seen in several parts of the HR-diagram. 
Associated properties of affected stars, such as age, mass, 
binarity, or evolutionary status,
do not appear to be uniform across the cases.  
Details in the observed abundance patterns seem to vary
between the cases, particularly for elements with
high T$_{cond}$, possibly due to the mechanism
itself or additional processes.
An accompanying infrared excess may be a warning bell, but it is 
not a required signature of dust-gas separation.
However, to date,  the inferred intrinsic metallicities 
of affected stars are one-tenth solar or greater. 

\section{Examination of dust-gas separation in the ultra metal-poor stars}

As the discussion turns to the ultra metal-poor stars, we  emphasize
the importance of the assumptions made in the determination of the $T_{cond}$ 
values.    Throughout our discussion, the $T_{cond}$ values are for a
solar composition gas.    Thus, the values are determined for a gas that 
is O-rich.   
One of the ultra metal-poor stars (HE0107-5240)
is presently severely C-rich, a second (HE1327-2326) has C/O $\simeq 1$,
and the C/O ratio is unknown for the third (HE0557-4840). 
Condensation of grains from C-rich gas provides C-rich solids 
(e.g., graphite and carbides) and a different set of condensation
temperatures. 

Calculations by Lodders \& Fegley (1995, 1999) have discussed the
condensation of grains in C-rich gas. 
For a gas with C/O $>$4, 
at a pressure  representative of a circumstellar region in
which dust forms via equilibrium chemistry (P$\sim 10^{-7}$ bar), 
the condensation temperatures $T_{cond}$ for graphite, TiC, and SiC 
(the first three condensates) 
are approximately 2000, 1560, and 1390 K, respectively. 
To lower temperatures, the condensation sequence is Fe, AlN, and CaS.

However, the $T_{cond}$'s are not the whole story. While Ti is 
effectively removed from the gas at temperatures below its $T_{cond}$, Si's
removal is constrained  by the fact that gaseous SiS is
resistant to condensation. The $T_{cond}$ for graphite but not for
TiC and SiC declines as C/O is decreased to near unity.
Condensation temperatures for Al, Mg,  Ca,  and
Fe are all several hundred degrees less than that of Ti. 
Thus, if the effective
$T_{cond}$ in the grain forming region is 1500~K or so, dust-gas separation
in a C-rich environment
should provide a pronounced Ti deficiency and only for much lower
temperatures will deficiencies for other elements be appreciable.
Finally, it is possible that dust-gas separation may remove sufficient 
carbon as graphite  
to lower the C/O ratio towards unity
 (Lodders \& Fegley 1999).    

In our figures, we adopt the T$_{cond}$ values determined for a 
solar (O-rich) composition gas, but discuss possible deviations due to the 
C/O ratio.

\subsection{Has dust-gas separation modified the composition of
HE1327-2326?}

Our discussion of HE1327-2326's composition is based on the 
abundance analysis by Frebel \& Christlieb (2007), which is drawn from new, 
high signal-to-noise VLT/UVES spectra, but confirms and extends previous
detailed analyses (Aoki \etal 2006; Frebel \etal 2006).  The abundances
are plotted as [$X$/H] versus $T_{cond}$ in Figures~2 \& 3, 
and assume the star is a subgiant
(\teff = 6180~K, \logg = 3.7; though our conclusions would be the same,
and the following discussion negligibly affected, were we to adopt the abundances 
assuming the star to be a dwarf).
While a classical model atmospheres LTE analysis  is performed,
predicted corrections are included for 
effects of stellar granulation.  We also show abundances 
[$X$/H] for two extreme post-AGB stars (HR~4049 and HD~52961), the 
RV Tauri variables (HP~Lyr and UY~CMa), 
and two Lambda Bootis  stars (29 Cyg and HD 106223). 

There is  a resemblance between the compositions of
HE1327-2326 and the chemically peculiar stars; 
as T$_{cond}$ increases, the elemental depletion increases
relative to solar.
However, HE1327-2326 is unique because of its extreme [Mg/Fe] 
ratio (see Figure 2) which is not seen in any of the chemically
peculiar stars.  Iron appears to be an outlier, since it
is the most underabundant element in this star, which is rarely 
the case for the RV Tauri or Lambda Bootis stars. 
Notably, S and Zn, important elements of low condensation temperature,
 have not been detected in  HE1327-2326.  
The S and Zn upper-limits provide no useful constraints
on HE1327-2326, although 
an abundance estimate for S may be possible in the future 
(see below).

Were we to insist that the abundance pattern for the RV Tauri 
star HP~Lyr, with its smooth trend in [$X$/H] vs T$_{cond}$, 
be a fair template for testing the suggestion that dust-gas 
separation has affected HE1327-2326, we would be bound to 
note the scatter in Figure~2 for $T_{cond} \geq
1200$ K. In particular, the low abundance of Fe relative to
Al and Ti in HE1327-2326 is the opposite in HP~Lyr.   
It seems unlikely that a change in the applied 3D corrections 
for the (as yet) untested models of stellar granulation can 
reverse this trend. The 3D
corrections are small for these elements and all of the same sign 
with an element-to-element scatter of less than 0.2 dex. 
Estimates of non-LTE corrections
assembled by Aoki \etal (2006) for the 1D model 
also aggravate the situation in that the corrections raise 
the Al abundance by 0.4 dex relative to Fe. 
The non-LTE corrections applicable to the 3D model are unknown.   

HP Lyr (and the other RV Tauri comparison star, UY CMa,
in Figure~2) may be an imperfect template for HE1327-2326.
Perhaps, {\it crucially}, the $T_{cond}$
estimates are based on a solar composition and, in particular,
on the fact that the Sun is O-rich (i.e., O/C $>1$). The measured
abundances of C and O  show that HE1327-2326 is nominally
C-rich  now as a subgiant and most likely the C/O ratio has been
depressed in evolution to the subgiant branch. (The C abundance is a mere
0.06 dex greater than the O abundance, a difference less than the
errors of measurement.) 
However, an alternative scenario, dust-gas separation in a C-rich environment 
also appears to fail to explain the observed composition.
As noted above, a signature of such a separation should be an
appreciable underabundance of Ti, which is not seen in Figure~2
relative to the other elements.

A comparison to the Lambda Bootis stars is more intriguing.   
Other than the large [Mg/Fe] ratio already noted as peculiar to
HE1327-2326, the relative abundances of Mg, Ca, and Ti to each
other (all high T$_{cond}$ elements) are similar to the Lambda Bootis 
 star HD106223 (see Figure~3).
If the mechanism for dust-gas 
separation is more similar between HE1327-2326 and the Lambda Bootis
stars, possibly complicated by diffusion or another process (see Section 2),
then these stars may be a better comparison template.
In Figure~3, both of the Lambda Bootis stars shown have similar 
depletions of Ca, Ti, and Fe.  Unfortunately, Al is not available
for either Lambda Bootis star.
Al, with its very high T$_{cond}$ is 
severely depleted in the RV Tauri variables, but less so in 
HE1327-2326.


If HE1327-2326 has undergone dust-gas separation, we can
estimate the star's intrinsic metallicity from the C, N, and O
abundances:
an initial metallicity of about $-2.0$ by the 3D-corrected C, N, and O
abundances but about $-1.3$ without these corrections.
The 3D corrections are large ($\simeq-0.7$ dex) for C, N, and O because
their abundances are derived from molecules (CH, NH, and OH) whose
formation is greatly enhanced by the presence of cooler regions
in 3D models. At an initial metallicity of about $-1.3$, the star resembles
some post-AGB  and RV Tauri stars.

If the initial metallicity of HE1327-2326
 were in fact either [Fe/H] = $-2.0$ or $-1.3$,
this has a small impact on our expectation of the initial abundance 
ratios for the other elements.  For example, [Ca/Fe] $\simeq +0.3$ for 
{\it normal} metal-poor stars with [Fe/H] = $-2$, and this is
approximately seen in Figure~2.   
A similar ratio  can be expected of Mg and Ti.  
While [Ti/Fe] is slightly larger than expected, the uniquely high 
[Mg/Fe] ($\simeq +2$)
 ratio is quite out of line with expectations 
for normal metal-poor stars.
Lastly, in normal metal-poor stars, [Al/Fe] $<0$,
thus the high Al point is also peculiar to HE1327-2326.

Inspection of the [$X$/H] for HE1327-2326 (and the other two stars) shows
no evidence for a relationship between [$X$/H] and the FIP.

\subsection{Has dust-gas separation modified the composition of
HE0107-5240?}

The composition of HE0107-5240 was determined by Christlieb \etal
(2004) and Bessell \etal (2002) who found the star to be
a giant with $T_{\rm eff} = 5100$ K and \logg =2.2. Their
analyses adopted a classical (1D) atmosphere and LTE for the
atmosphere and line formation. The derived abundances  corrected
to those for a 3D atmosphere by Collet \etal (2006) with retention
of the  assumption of LTE  are plotted in Figures~4 \& 5.  

The abundance pattern for HE0107-5240 is more similar to the chemically
peculiar stars than was HE1327-2326 discussed above.  The [Mg/Fe] ratio
is similar to that of the chemically peculiar stars, as well as normal
metal-poor stars.   Ca and Ti also show depletions with respect to Fe
that are in fair agreement with those of the RV Tauri stars.   
The upper-limits to Zn and S
provide no useful constraint for examining the dust-gas scenario, 
however the very low upper-limit for Al is consistent with the
predictions.

This star is presently C-rich; C/O (by number) 
is about 20 for the 1D analysis and about 6 for the 3D analysis. 
An initial metallicity of $\simeq -4$ is crudely indicated by the 3D 
abundances of N and O, but $\simeq -2$  by the 1D C abundance.   
The result of the 1D to 3D corrections is, as indicated above, a large
reduction in the C, N, and O abundances, all derived from
diatomic molecules (CH, C$_2$, NH, CN, and OH). In 1D, the
abundances are larger than those plotted by about 1.2, 1.1, and 0.7
dex for C, N, and O, respectively.  It is an interesting point, possibly
one of concern, that NH and CN (assuming the C abundance
from CH and C$_2$) give a consistent N abundance using the 1D
atmosphere but inconsistent abundances (difference is 0.9 dex)
using the 3D atmosphere. 
The [C/O] ratio in HE0107-5240 
is similar to the carbon-enhanced metal-poor stars
which also range in metallicity from [Fe/H] = $-2$ to $-4$
(Sivarani \etal 2006).

The appropriateness of using $T_{cond}$ in Figures~4 \& 5 is decidedly
questionable due to the star being very C-rich. 
If the star was and is a single star and its evolution  from the
main sequence to its present status as a giant even approximately
follows expectation, the star must have begun life even more C-rich. 
Then, if dust-gas separation occurs in either a protostellar, 
circumstellar or wind environment, it is the $T_{cond}$'s for 
C-rich gas that  are the relevant quantities. 
As noted above, a significant deficiency of Ti is
expected for the star accreting substantial amounts of gas.  
On the other hand, if the star has accreted significant
amounts of material from the interstellar medium of the established Galaxy,
including the present diffuse interstellar medium, the dust-gas separation
was likely to have occured or is occuring in an O-rich medium and,
then Figures~4 \& 5 use approximately the appropriate $T_{cond}$ values. 
Accretion of interstellar gas that is O-rich requires then an explanation
of how HE0107-5240 is so C-rich and metal-poor.

\subsection{Has dust-gas separation modified the composition of
HE0557-4840?}

The composition is taken from Norris \etal (2007) who determine
that the star is a giant and offer  spectroscopic and photometric
estimates of the effective temperature of 5100 K and 4900 K,
respectively. The surface gravity is given as \logg = 2.2. 
The mean of their LTE abundances for a 1D atmosphere are plotted
in Figure~6. Corrections for adoption of a  3D model atmosphere
are small, except for C (and the upper-limits on N and O), 
and are plotted in Figure~7. 
The C/O ratio is undetermined for this star.

If the initial composition were that suggested by the C abundance 
and N upper-limit for a 1D atmosphere, the initial [Fe/H] $\simeq -3$.  
Taking the 3D corrections into account,  
the initial metallicity is likely $\simeq -4$.
This lower initial abundance also reduces the scatter in the
[$X$/H] values, with a hint that [$X$/H] is independent of $T_{cond}$.

There is no pattern in the elemental abundances for HE0557-4840 that
would clearly suggest dust-gas separation.   The [$X$/H] values for nearly
all elements are flat, ranging from $-4.0 > [$X$/H] > -5.5$, and
therefore showing similar depletions from a solar composition gas
in nearly all elements.  
This might be unexpected since the normal metal-poor stars 
with [Fe/H] $\sim -4$ show higher $\alpha$-element abundance ratios, 
and usually an extremely wide range in the $r$-process elements
(McWilliam 1997).   However, the [$\alpha$/Fe]
abundances in HE0557-4840 do not fall outside  the range of normal
metal-poor star abundances (e.g., [Mg/Fe] and [Ti/Fe] are
within expectations).

%
%
%

\subsection{Other Metal-Poor Stars with [Fe/H] $\le -3.5$ }

It is natural to ask how ubiquitous the dust-gas separation pattern
is amongst the extremely metal-poor stars with detailed elemental
abundance determinations.    An examination of the literature for
stars with $-3.5 \ge$ [Fe/H] $\ge -4.0$ shows that none of them 
shows the dust-gas separation signature.   Some examples:
CS 29498-043 (Aoki \etal 2004) shows an enhancement in Mg similar
to those for CNO, and the Zn upper-limit is significantly lower 
than CNO which is not predicted in the dust-gas separation pattern. 
CS22949-037 (Depagne \etal 2002) also shows an enhancement in Mg
(and Na) that is close to that for CNO, and the Zn determination 
is signficantly lower whereas dust-gas separation would predict a
very similar Zn enhancement to those other elements. 
Finally, for HE1300+0157 (Frebel \etal 2007) and HE1424-0241 
(Cohen \etal 2007), several elements ($\alpha$ elements and iron
group elements) have the same enhancements (or nearly so) as the 
CNO abundances.
Of course, each of these stars has interesting and unusual 
abundance pattern(s), and while not similar to the predictions
for dust-gas separation, they are interesting  in terms of the 
progenitor mass or explosion characteristics 
of their (small number of) contributing supernovae.

\section{Infrared excesses}

An association of circumstellar dust, i.e., an infrared excess, with
a star that is a candidate for exposure to dust-gas separation
strengthens the explanation for the abundance anomalies.
The absence of an infrared excess is not fatal to the star's
candidacy, as shown by many of the Lambda Bootis stars, and the
post-AGB star BD+39 4926.

The broad-band colors and spectral energy distributions from 
model atmospheres are shown for the three of the ultra metal-poor
stars in Figures~8 to 10.   JHK magnitudes are from the 2MASS catalogue; 
UBVRI magnitudes are from Aoki \etal (2006) for HE1327-2326, 
Christlieb \etal (2002) for HE0107-5240, 
and Beers \etal (2007) for HE0557-4840.  UBVRI
magnitudes have been converted to fluxes using the Vega zero 
points by Colina \etal (1996) and JHK magnitudes converted
using the 2MASS Explanatory Supplement (Sect VI.4a).
Reddening has been taken from the reddening maps by 
Schlegel \etal (1998); E(B-V) = 0.08, 0.013, and 0.04
for HE1327-2326, HE0107-5240, and HE0557-4840, respectively. 
These were converted 
to A$_\lambda$ based on conversions in Cardelli \etal (1989).

The observed fluxes are compared to model flux distributions from 
the online grid of available MARCS (Gustafsson \etal 1975; 
Asplund \etal 1997) and Kurucz models (Castelli \& Kurucz 2003).
The closest model parameters to those adopted for each star
and available in both grids were 
\teff = 6250 K, \logg = 4.0, and [Fe/H]=$-5.0$ for HE1327-2326,
and  \teff = 5000 K, \logg = 2.0, and [Fe/H]=$-5.0$ for both
HE0107-5240 and HE0557-4840.  
There are small differences in the models that account for
their slight differences in spectral energy distributions,
e.g., there are differences in the opacities used, and 
MARCS models vary stellar mass and [$\alpha$/Fe] ratios. 
The observed fluxes were scaled
to overlay the model fluxes best at R, I, and J; the scale factor
is estimated for each star, 
but would be equivalent to (distance/radius)$^2$.

No infrared excess is seen in the colours of any of the ultra metal-poor
stars compared to the model flux distributions.    
While this is an important test
for the presence of a dusty disk, it is not definitive.   Emission
from a disk depends on composition, age, distance from the star,
and dust temperature.   For many post-AGB stars, like HR 4049,
the infrared excess is  strong in the K-band 
and accompanied by a suppression in the UV continuum 
(Figure~1 of Dominik \etal 2003).  
On the other hand, the circumstellar disk around the post-AGB 
star HD56126 (Van Winckel \& Reyniers 2000) is detected  only
beyond 4 microns, with no excess observed in the K-band or below 
(see Figure 1 in the review paper by van Winckel 2003).
Furthermore, not all Lambda Bootis stars show evidence of their
dusty disks either.  Paunzen \etal (2003)
estimated 23\% of {\it bona fide} Lambda Bootis stars show evidence for
circumstellar matter via infrared excesses or Infrared Space
Observatory (ISO) and sub-mm
CO(2-1) line emission.   Also, the presence of dust
around other A-type stars does not necessarily imply the presence
of the Lambda Bootis abundance pattern (Kamp \etal 2002, Dunkin
\etal 1997); in the study by Acke \& Waelkens (2004) only one
in 24 targets showed a clear Lambda Bootis abundance pattern.

Obviously, the search for an infrared excess should, if possible, be
extended to longer wavelengths.  
Although an infrared excess would clearly identify the presence of
a dusty disk, the absence of infrared emission does not preclude the
dust-gas separation scenario.

\subsection{Are these stars spectroscopic binaries?}

The presence of a companion seems a prerequisite for efficient
dust-gas separation in a post-AGB A-type star and quite probably
in the RV Tauri stars. (It is not seemingly a  prerequisite
for the Lambda Bootis stars.) For this and other reasons, 
it would be valuable to show whether all three stars are binaries. 
Presently, the available data are limited on this point.

For HE0107-5240, Bessell \etal (2004) find no radial velocity 
variations larger than 0.5 km~s$^{-1}$ over $\sim$100 
(or 373 days when including two spectra taken nearly a year earlier).
Similarly for HE1327-2326, Frebel \etal (2006) find no variations
larger than their measurement error or 0.7 to 1.0 km/s over 
$\sim$100 days (383 if 3 spectra taken nearly a year earlier
are included).  For HE0557-4840, Norris \etal (2007) find no velocity 
variations greater than $\pm0.4$ km s$^{-1}$ from observations spanning 
40 days.  These data do not find evidence for binarity, however
The gold standard for mass transfer across a binary creating 
abundance anomalies is McClure's (1985) study of velocity 
variations for Barium stars, which included a 5 year campaign,
with accuracies of 0.5 km s$^{-1}$. 
An intensive observing campaign may be needed to test convincingly 
the hypothesis that these ultra metal-poor stars are or are not
spectroscopic binaries.

\subsection{The sulphur abundance}

For HE1327-2326, but not for the cooler stars HE0107-5240 and HE0557-5240,
it may be possible to detect S\,{\sc i} lines and determine 
or at least set an interesting limit on the S abundance.
Observations of Lambda Bootis stars, RV Tauri variables, and A-type
post-AGB stars  show that S (and Zn) do not share the depletions of
iron-group elements (Figure 1). 
The upper-limit on the Zn abundance for HE1327-2326 
is consistent with a smooth interpolation of the
C, N, O, and Na abundance.

The strongest S~I lines of three different multiplets are 
observed  around 10455, 9212, and 8695 \AA\,
(Caffau \etal 2005, 2007). 
Given the atmospheric parameters of HE1327-2326,  
we have performed a spectrum synthesis using MOOG
(Sneden 1973) and OSMARCS models (Plez \etal 2000).
If the star has undergone dust-gas separation, then
the intrinsic metallicity as suggested by the C, N, and O
abundance is [$X$/H] $\simeq -2$ according to the 3D models but
$-1.3$ for 1D models.
At [$X$/H]=$-2$, the strongest S\,{\sc i} lines of the
10455, 9212, and 8695 \AA\ multiplets, 
are estimated to have LTE equivalent widths of 
15, 25, and 1 m\AA.   NLTE corrections are thought to strengthen
the line from these predictions (Takeda \etal 2005).
Thus, the 10455 \AA\, and 9212 \AA\, lines should be  detectable,
however the 9212 \AA\, lines are  in a region of strong 
telluric lines.

Detection of the 10455 \AA\ triplet was reported by
Nissen \etal (2007). These authors used CRIRES on the VLT to detect
with ease the S\,{\sc i} lines in G29-23, a subdwarf with [Fe/H] $= -1.7$.
Its atmospheric parameters ($T_{\rm eff} = 6200$ K, and \logg = 4.0)
are similar to those of HE1327-2326. Its [Fe/H] is within the range
supposed for HE1327-2326 were the C, N, and O abundances indicative
of the initial composition. The challenge will be to
extend the CRIRES observation of G29-23 at V = 10.2 to HE1327-2326
at V = 13.5, a decrease of about a factor of 20 in flux.

\section{Concluding remarks} 

In this paper, we have outlined the existing evicence for and against
the three ultra metal-poor stars being affected by dust-gas separation.
The evidence for dust-gas separation having occured at the stellar 
photosphere is primarily the abundance pattern, which resembles that
of known chemically peculiar stars globally where dust-gas separation
has been supported by other observations.   The other observations 
usually include one or more of the following for at least one object
within a sample; infrared excesses (usually beyond the K band at 2.2 microns) 
or other evidence of circumstellar material, evidence of binarity, or
the observation that sulphur or zinc do not have the same depletions 
as the rest of the metals.   The latter is a critical test because
non-depleted abundances of these two elements is more consistent with 
dust formation, otherwise requiring random and puzzling variations in 
nucleosynthesic yields.    Most of these observational tests do not
yet exist for the ultra metal-poor stars, or the results are currently
inconclusive regarding dust-gas separation.   

Searchs for radial velocity variations are ongoing in the ultra metal-poor
stars to determine if they are in binary systems.  Currently, no variations
are found, which suggests these stars are not binaries, but this evidence
cannot be conclusive because of orbit inclinations and/or potentially small
amplitude variations.     Sulphur and zinc upper-limits exist for all three
ultra metal-poor stars, however these upper-limits are presently too high 
to distinguish between nucleosynthetic origins or dust-gas separation.  
K band photometry exists for all three ultra metal-poor stars, and none of
them show an infrared excess, however many RV Tau, post-AGB, and Lambda Boo
chemically peculiar stars only show infrared excesses beyond the K band, if at
all (many have no detectable infrared excesses).    
In addition, known chemically
peculiar stars suggest that neither dust not binarity are
necessary conditions; the Lambda Bootis stars  exist as single
stars, often without dust; the peculiar post-AGB are binaries but
one lacks dust; the affected RV Tauri stars have an infrared
excess and, although binarity may be a necessary condition, it is
as yet unknown observationally that all are binaries.
Nevertheless, these observational are possible for the ultra metal-poor 
stars, 
e.g.  Spitzer observations at mid-IR wavelengths, and examination 
of the S\,{\sc I} lines in near-IR spectroscopy,
but have yet to be performed.

One remaining concern is that two of the ultra metal-poor stars do not 
occupy similar atmospheric parameter ranges as any of the known
chemically peculiar stars. 
HE0107-5240 and HE0557-4840 have temperatures like the RV~Tauri
variables, yet higher gravities and lower intrinsic metallicities
(if CNO are the appropriate proxies). 
Physically, these stars are expected to have atmospheres  with deep
convective envelopes, and thus, accreted gas from a stellar wind, 
a circumstellar shell, or a circumbinary disk is should be diluted 
beyond detection.    
Existence of these envelopes makes it difficult to understand how 
dust-gas separation effects can be created and sustained in these giants.
Indeed, the stars must have received several tenths of a solar mass of 
separated gas in order that the anomalies be detectable in these giants.
Of course, this is a concern for the RV Tauri and post-AGB stars as well.
These difficulties are  less severe for HE1327-2326, which is 
significantly hotter and is expected to have a more shallow
convective envelope.   HE1327-2326 occupies (nearly) the same 
stellar parameter range as the Lambda Bootis stars.  Fortunately, 
detection and analysis of the S\,{\sc i} lines in this star offers 
a critical test for any chemical peculiarities due to dust-gas separation.

Stellar astronomers are conditioned to think about stars with
very peculiar abundance anomalies - real or imagined. Commonsense
is often a reliable guide to the boundary between real and imagined.
In this regard, if the three stars under consideration here
are truly stars of a higher intrinsic metallicity, one
might ask why examples have not been seen in well studied
samples of metal-poor stars. The globular clusters spring to
mind. Str\"{o}mgren photometry should be able to pick out those
few stars
with a reduced metallicity from the rest of the stars showing
the monometallicity that is a mark of a globular cluster.
Of course, if these stars are due to the random effects of
passing through dense interstellar clouds, and the effects on
the abundances are short lived, then one does not expect to find
similar stars in globular clusters, and from objective prism surveys
such stars would not stand out as peculiar if the resultant metallicity
is greater than $\sim -3$. 

  It is only because ultra metal-poor stars
are so important as tests of early chemical evolution in the Galaxy
that these stars were picked out and studied in detail from high
resolution spectroscopy.   It is 
important that the suspicion be laid to rest that
they are not truly ultra metal-poor but
chemically peculiar stars  of a more common, if low,  metallicity. 
Then, the focus may be placed exclusively on finding an explanation
in terms of stellar nucleosynthesis and the chemical evolution of
the young Galaxy (e.g., Iwamoto \etal 2005).

\acknowledgments
We are grateful to P. Bonifacio for valuable discussions on
metal-poor stars and on observations and model calculations 
of the S\,{\sc I} lines.
We thank Anna Frebel for providing abundances in advance of
publication and Katharina Lodders for helpful comments on dust formation
in C-rich material. 
Thanks to Ian Roederer, Inga Kamp, and the anonymous referee for 
many helpful comments on this manuscript.
DLL's contributions have been supported by the
Robert A. Welch Foundation of Houston, Texas.   
KAV would like to thank NSERC for support through a Discovery grant.

\begin{figure}
\plotone{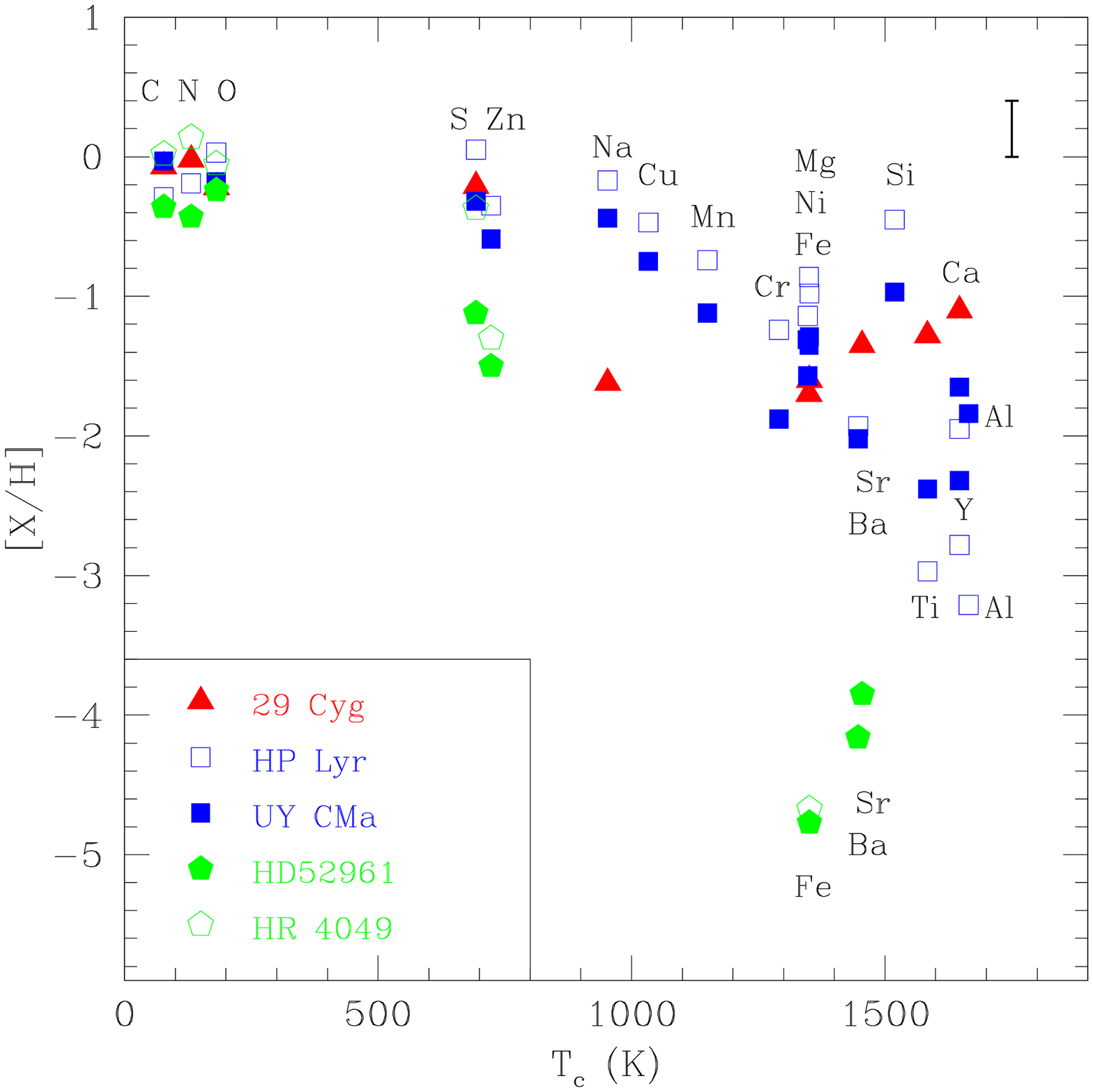}
\caption{ 
Elemental abundances of chemically peculiar stars relative to solar; 
including the Lambda Bootis star, 29~Cyg (red; Venn \& Lambert 1990), 
two RV Tauri stars 
UY~CMa and HP~Lyr (Giridhar \etal 2005), and two A-type post-AGB stars,
HR~4049 and HD~52961 (Waelkens \etal 1991, with Zn from 
Takeda \etal 2002 and van Winckel \etal 1992).    
Elemental abundances are plotted 
against the solar condensation temperatures from Lodders (2003).
Typical errors are $\le0.2$~dex, shown by the errorbar in the upper right.
}
\end{figure}

\begin{figure}
\plotone{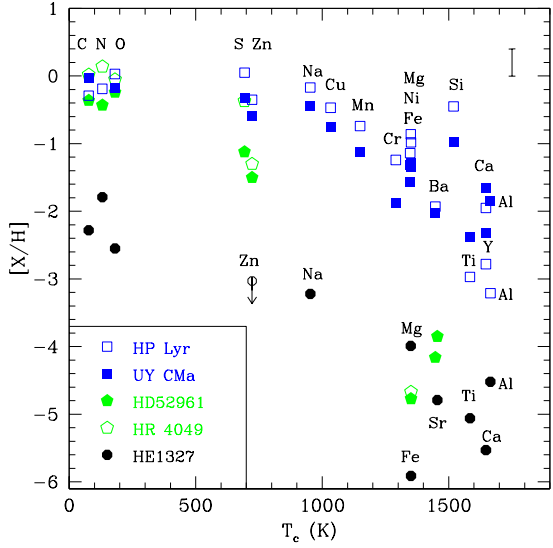}
\caption{ 
[$X$/H] abundances in HE1327-2326 from Frebel \& Christlieb (2007)
against the solar condensation temperatures from Lodders (2003).
LTE abundances (solid points) and the upper limit for Zn are shown, 
all with (3D) corrections for stellar granulation.  
This abundance pattern is compared to the post-AGB and RV Tauri
stars (references in Figure~1).   While the abundance pattern
for HE1327-2326 resembles those of the chemically peculiar stars,
the high [Mg/Fe] ratio and Al values are striking. 
Typical errors are $\le0.20$~dex shown by the errorbar in the upper right;
Aoki \etal (2006) suggest slightly larger errors for C and N in HE1327-2326,
of 0.24 and 0.30 dex, respectively. 
}
\end{figure}

\begin{figure}
\plotone{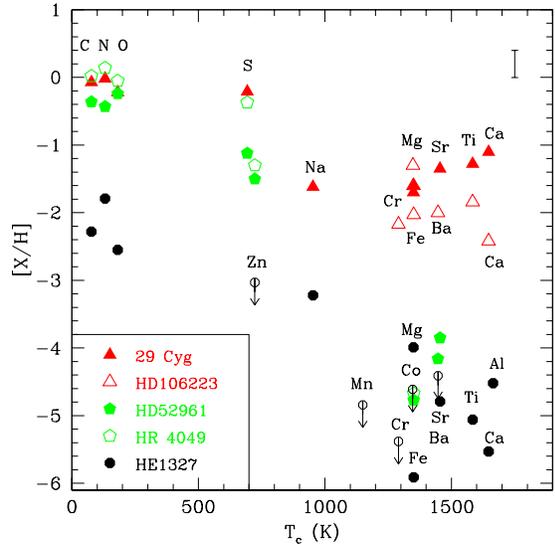}
\caption{ 
[$X$/H] abundances in HE1327-2326 from Frebel \& Christlieb (2007)
as in Figure~2 with additional upper-limits noted.
The abundance pattern is compared to the post-AGB stars 
and two Lambda Bootis stars (29 Cyg from Venn \& Lambert 1990, 
and HD~106223 from Andrievsky \etal 2002).   
}
\end{figure}

\clearpage

\begin{figure}
\plotone{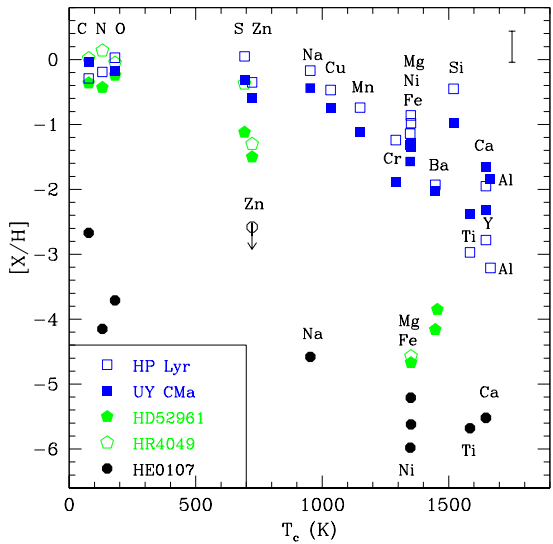}
\caption{ 
[$X$/H] abundances in HE0107-5240 from Collet \etal (2006;
originally from Christlieb \etal 2004 and Bessell \etal 2004) 
are shown against the solar condensation temperatures from 
Lodders (2003).
LTE abundances (solid points) and the Zn upper-limit are shown, 
all with (3D) corrections for stellar granulation.
This abundance pattern is compared to the post-AGB stars 
and RV Tauri variables as in Figure~2.
Typical errors for HE0107-5240 are $\le0.24$~dex, 
shown by the errorbar in the upper right;
only the uncertainly on C could be higher, 0.34~dex when 
determined from the CH {\it A-X} band (Christlieb \etal 2004). 
}
\end{figure}

\begin{figure}
\plotone{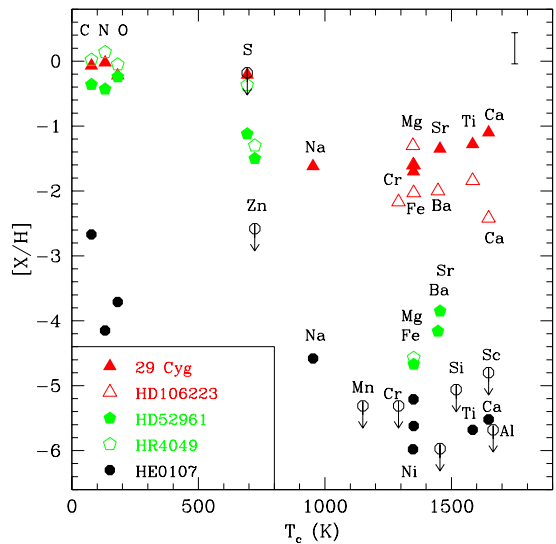}
\caption{ 
[$X$/H] abundances in HE0107-5420 from Collet \etal (2006)
as in Figure~4 with additional upper-limits shown. 
The abundance pattern is compared to the post-AGB stars 
and Lambda Bootis stars as in Figure~3.
}
\end{figure}

\clearpage

\begin{figure}
\plotone{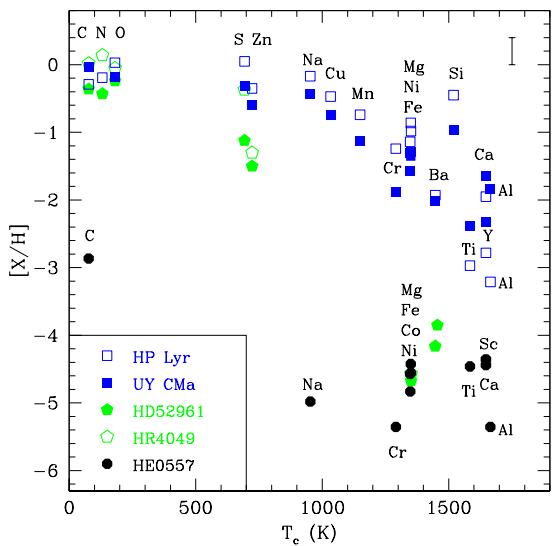}
\caption{ 
[$X$/H] abundances in HE0557-4840 from Norris \etal (2007) are shown
against the solar condensation temperatures from Lodders (2003).
LTE abundances (solid points) are shown with {\it no corrections
for 3D effects}. 
This abundance pattern is compared to the post-AGB stars 
and RV Tauri variables as in Figure~2.
Typical errors are $\le0.20$~dex, shown by the errorbar in the upper right.
}
\end{figure}

\begin{figure}
\plotone{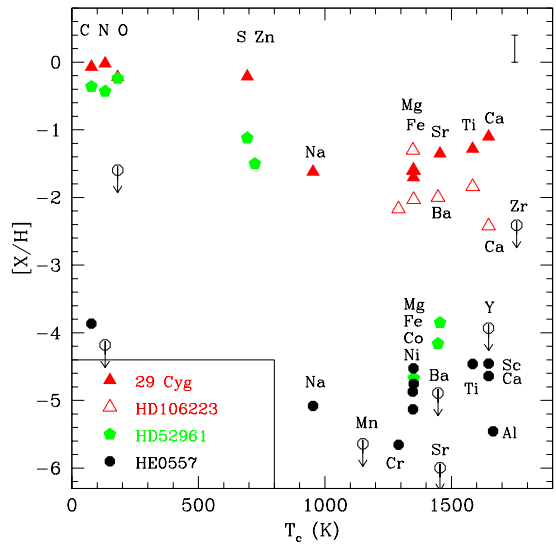}
\caption{ 
[$X$/H] abundances in HE0557-4840 from Norris \etal (2007) are shown
against the solar condensation temperatures from Lodders (2003).
LTE abundances (solid points) that have been corrected for
solar granulation (3D effects) are shown.  The 3D corrections 
have a very small effect on most elements, except carbon.
Typical errors are $\le0.20$~dex, shown by the errorbar in the upper right.
The abundance pattern is compared to one of the post AGB
stars and the Lambda Bootis stars as in Figure~3.
}
\end{figure}

\begin{figure}
\plotone{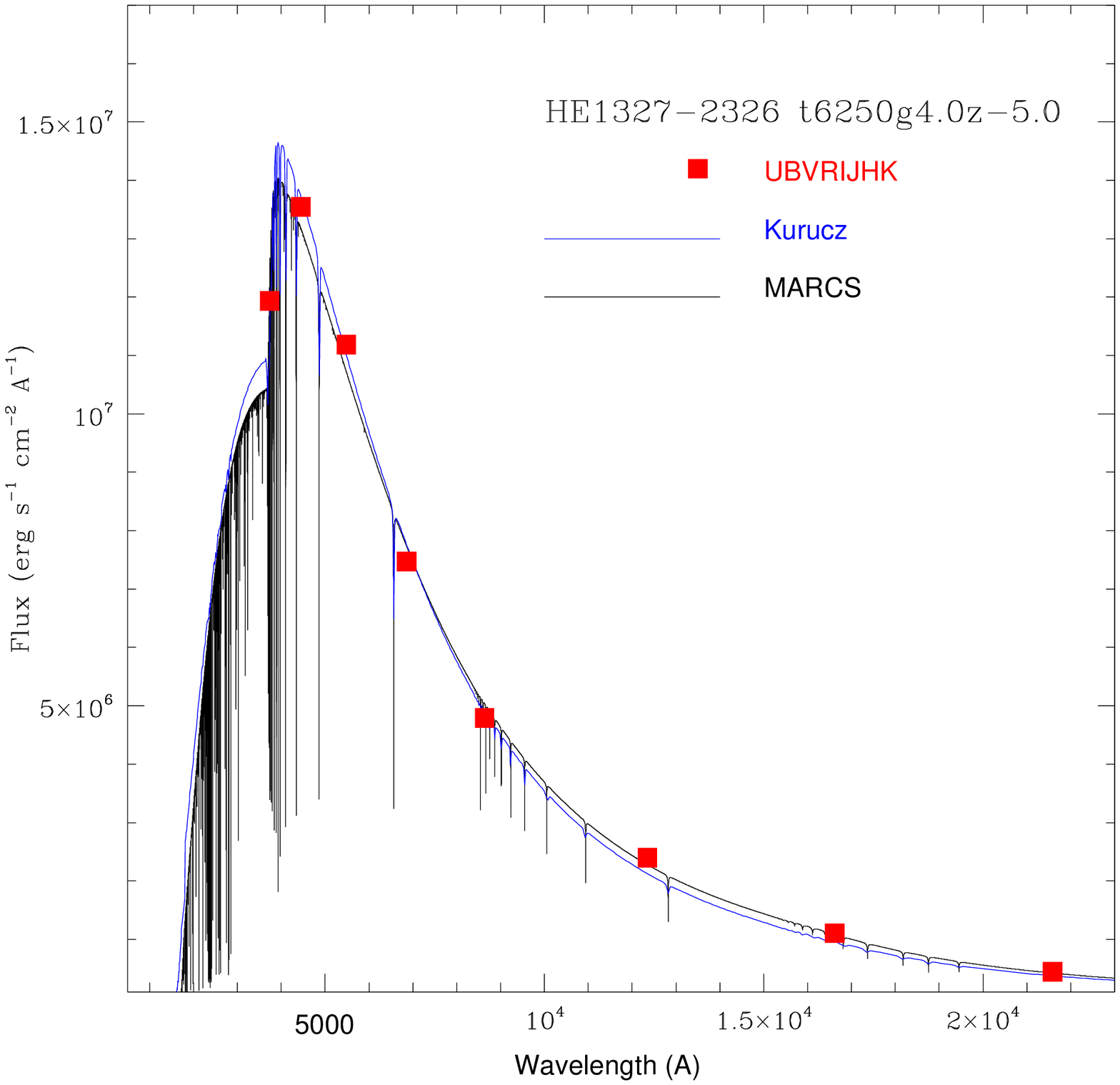}
\caption{
The spectral energy distribution for HE1327-2326 from broad band
UBVRI (Aoki \etal 2006) and JHK (2MASS) fluxes.   
These are compared to model atmosphere 
flux distributions from both MARCS and Kurucz with
T$_{\rm eff}$=6250\,K, \logg=4.0, and [Fe/H]=$-5.0$.   
Observed magnitudes are scaled to the absolute flux units
in order to best fit the RIJ magnitudes.
Errors in the fluxes are smaller than the symbol sizes.
}
\end{figure}

\begin{figure}
\plotone{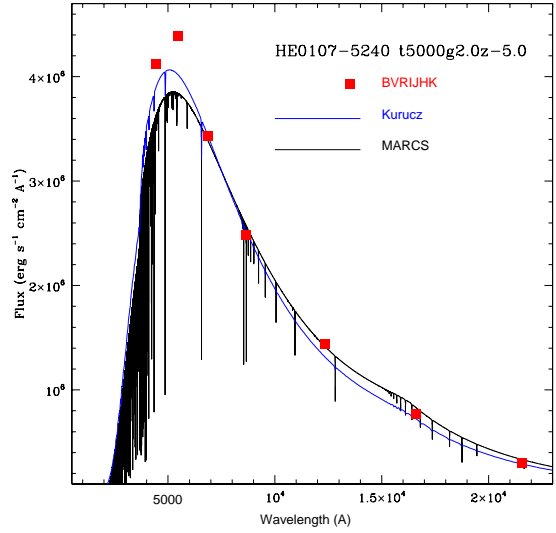}
\caption{
The spectral energy distribution for HE0107-5240.
Broad band BVRI colors from Christleib \etal (2002) and 
JHK (2MASS) colors.   The model atmosphere 
flux distributions are from both MARCS and Kurucz with
T$_{\rm eff}$=5000\,K, \logg=2.0, and [Fe/H]=$-5.0$.
Observed magnitudes are scaled to the absolute flux units
in order to best fit the RIJ magnitudes.
Errors in the fluxes are smaller than the symbol sizes.
}
\end{figure}

\begin{figure}
\plotone{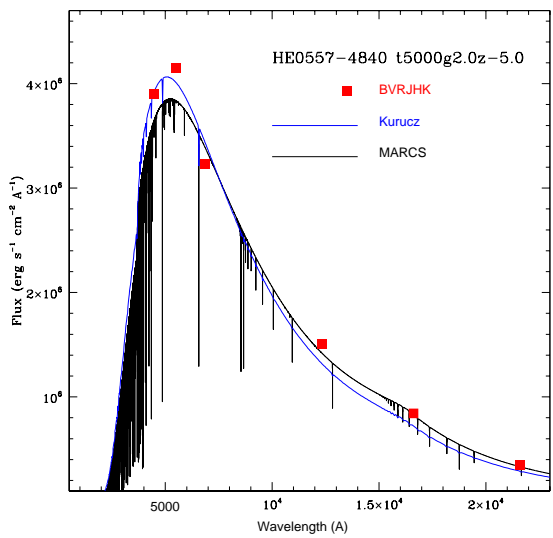}
\caption{
The spectral energy distribution for HE0557-4840 from broad band
BVR (Beers \etal 2007) and JHK (2MASS) fluxes.   
These are compared to model atmosphere 
flux distributions from both MARCS and Kurucz with
T$_{\rm eff}$=5000\,K, \logg=2.0, and [Fe/H]=$-5.0$.   
Observed magnitudes are scaled to the absolute flux units
in order to best fit the R and J magnitudes.
Errors in the fluxes are smaller than the symbol sizes.
}
\end{figure}


\end{document}